\documentclass[aps, prx, reprint, superscriptaddress, amsmath, amssymb, floatfix, longbibliography]{revtex4-2}

\usepackage{physics}% for braket julin
\usepackage[normalem]{ulem}
\usepackage{graphicx}% Include figure files
\graphicspath{ {pdffig/}, {supfig/}, {images/}, {figs/} }
\usepackage{dcolumn}% Align table columns on decimal point
\usepackage{bm}% bold math
\usepackage{hyperref}% add hypertext capabilities
\usepackage{siunitx} % For SI unit
\usepackage{ulem}    % For \sout{text}
\usepackage{pifont}
\usepackage{xcolor}
\hypersetup{
	colorlinks,
	linkcolor={blue!90!black},
	citecolor={blue!90!black},
	urlcolor={blue!90!black}  
}
\usepackage[caption=false]{subfig}
\makeatletter
\newcommand{\colorcaption}[2][]{%
  \begingroup%
  \renewcommand{\@caption@fignum@sep}{ (color online). }%
  \caption[#1]{#2}%
  \endgroup%
}
\makeatother

\def\Rb{$^{87}$Rb}
\def\Na{$^{23}$Na}

\def\NaRb{$^{23}$Na$^{87}$Rb }

\begin{document}

\title{Microwave shielding of bosonic NaRb molecules}

\author{Junyu Lin}
\affiliation{Department of Physics, The Chinese University of Hong Kong, Shatin, Hong Kong, China}

\author{Guanghua Chen}
\affiliation{Department of Physics, The Chinese University of Hong Kong, Shatin, Hong Kong, China}

\author{Mucan Jin}
\affiliation{Department of Physics, The Chinese University of Hong Kong, Shatin, Hong Kong, China}

\author{Zhaopeng Shi}
\affiliation{Department of Physics, The Chinese University of Hong Kong, Shatin, Hong Kong, China}

\author{Fulin Deng}
\affiliation{Key Laboratory of Artificial Micro- and Nano-structures of Ministry of Education, School of Physics and Technology, Wuhan University, Wuhan, Hubei 430072, China}

\author{Wenxian Zhang}
\affiliation{Key Laboratory of Artificial Micro- and Nano-structures of Ministry of Education, School of Physics and Technology, Wuhan University, Wuhan, Hubei 430072, China}

\author{Goulven Qu{\'e}m{\'e}ner}
\affiliation{Universit\'{e} Paris-Saclay, CNRS, Laboratoire Aim\'{e} Cotton, 91405, Orsay, France}

\author{Tao Shi}
\affiliation{CAS Key Laboratory of Theoretical Physics, Institute of Theoretical Physics, Chinese Academy of Sciences, Beijing 100190, China}
\affiliation{CAS Center for Excellence in Topological Quantum Computation \& School of Physical Sciences, University of Chinese Academy of Sciences, Beijing 100049, China}
\affiliation{Peng Huanwu Collaborative Center for Research and Education, Beihang University, Beijing 100191, China}

\author{Su Yi}
\affiliation{CAS Key Laboratory of Theoretical Physics, Institute of Theoretical Physics, Chinese Academy of Sciences, Beijing 100190, China}
\affiliation{CAS Center for Excellence in Topological Quantum Computation \& School of Physical Sciences, University of Chinese Academy of Sciences, Beijing 100049, China}
\affiliation{Peng Huanwu Collaborative Center for Research and Education, Beihang University, Beijing 100191, China}

\author{Dajun Wang}
\email{djwang@cuhk.edu.hk}
\affiliation{Department of Physics, The Chinese University of Hong Kong, Shatin, Hong Kong, China}
\affiliation{The Chinese University of Hong Kong Shenzhen Research Institute, Shenzhen, China}

\date{\today}

\begin{abstract}

Recent years have witnessed tremendous progresses in creating and manipulating ground-state ultracold polar molecules. However, the two-body loss regardless of the chemical reactivities is still a hurdle for many future explorations. Here, we investigate the loss suppression of non-reactive bosonic $^{23}$Na$^{87}$Rb molecules with a circular polarized microwave blue-detuned to the rotational transition. We achieve suppression of the loss by two orders of magnitude with the lowest two-body loss rate coefficient reduced to $3\times10^{-12}~\rm{cm^3/s}$. Meanwhile, the elastic collision rate coefficient is increased to the $10^{-8}~\rm{cm^3/s}$ level. The large good-to-bad collision ratio has allowed us to carry out evaporative cooling of $^{23}$Na$^{87}$Rb with an efficiency of 1.7(2), increasing the phase-space density by a factor of 10. With further improvements, this technique holds great promises for creating a Bose-Einstein condensate of ultracold polar molecules. 

\end{abstract}

%\pacs{}
\maketitle

\section{Introduction} 

Ultracold polar molecules (UPMs) with strong and long-range electric dipole-dipole interactions (DDI) have long been considered as a very promising platform for quantum science and technology~\cite{carr2009cold, baranov2012condensed,bohn2017cold}. For many years the progress in this research direction was hindered by the difficulty in creating samples of UPMs. Following the recent breakthroughs in both heteronuclear atom association~\cite{ospelkaus2008efficient, ni2008high, takekoshi2014ultracold, molony2014creation, park2015ultracold, guo2016creation, rvachov2017long, seeselberg2018modeling, voges2020ultracold, stevenson2023ultracold} and direct laser cooling of molecules~\cite{barry2014magneto, anderegg2017radio, truppe2017molecules,collopy20183D, ding2020sub}, this problem has now been largely solved. Currently, the main issue is the rapid two-body losses even without chemical reactions due the formation of two-molecule complexes~\cite{Mayle2013,ye2018collisions,gregory2019sticky,christianen2019photoinduced,hu2019direct,gregory2020loss,gersema2021probing,bause2021collisions}. This poses as a great challenge for achieving quantum degeneracy which is required as the initial condition for many applications. Several methods have been investigated, including the optical lattice isolation~\cite{chotia2012long, yan2013observation, reichsollner2017quantum, lin2022seconds}, 2D suppression with in-plane DDI~\cite{quemener2011dynamics, de2011controlling}, and engineered long-range interaction potential barriers via DC field induced F\"{o}rster resonance~\cite{avdeenkov2006suppression, matsuda2020resonant, li2021tuning} and blue-detuned microwave shielding~\cite{karman2018microwave, lassabliere2018controlling, karman2020microwave, anderegg2021observation, schindewolf2022evaporation} or optical shielding~\cite{xie2020optical,karam2022two}, to mitigate this issue. With the latter methods, quantum degenerate Fermi gases of $^{40}$K$^{87}$Rb and $^{23}$Na$^{40}$K molecules have been produced very recently~\cite{valtolina2020dipolar, schindewolf2022evaporation}. However, for bosonic UPMs, which suffer from the loss more severely due to the absence of the $p$-wave barrier, Bose-Einstein condensates have not been realized.

\begin{figure*}[t]
    \centering
    \includegraphics[width=0.75\linewidth]{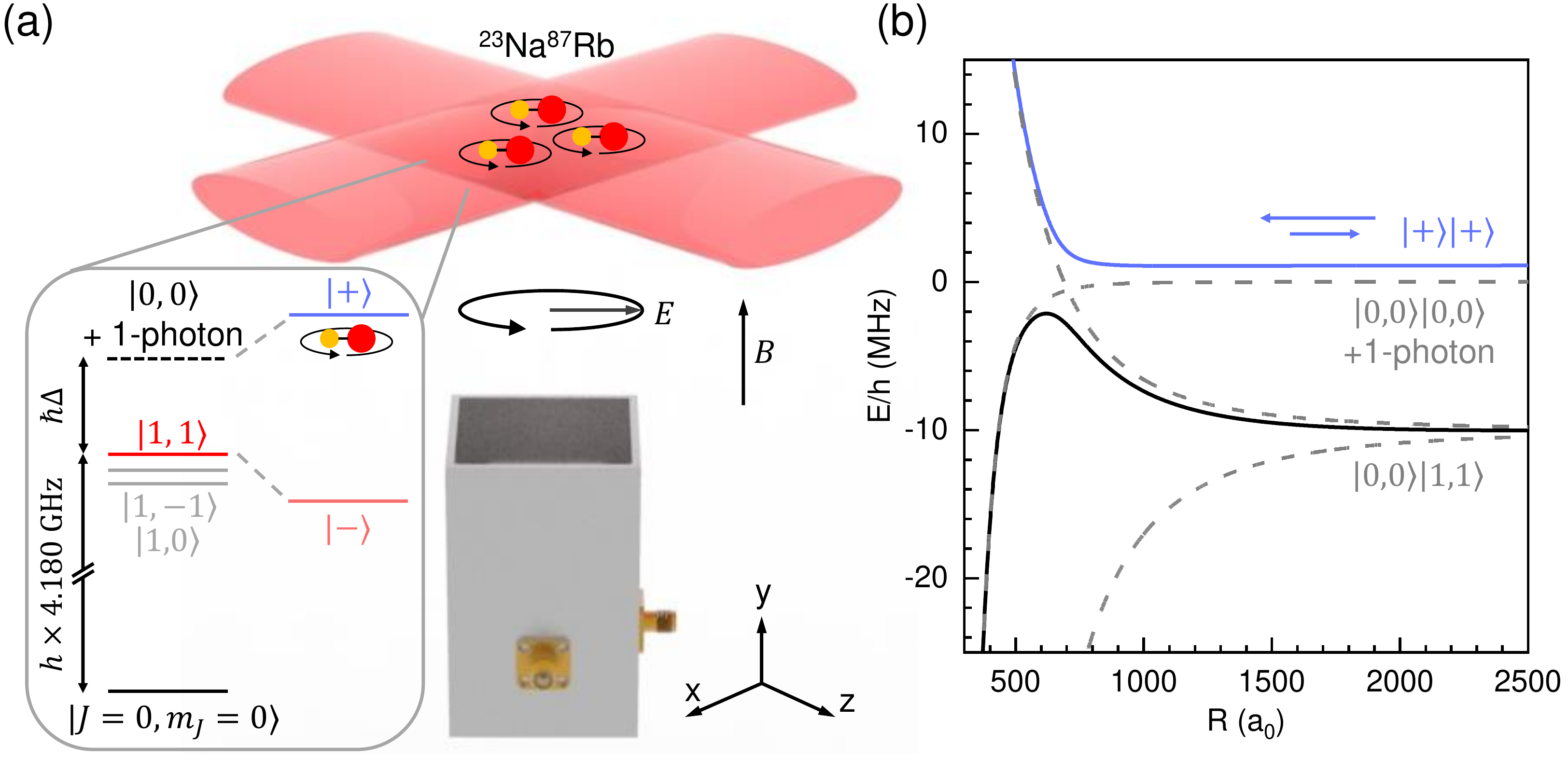}
\caption{Microwave shielding of \NaRb molecules.
	(a) Sketch of the experiment setup and the related energy levels. The optical dipole trap is formed by crossing two elliptical shape 1064 nm laser beams. A 346 G magnetic field in the vertical direction always presents. The blue-detuned and circular polarized microwave field is broadcast to the sample through a doubly-fed waveguide, coupling the $\ket{0,0} \leftrightarrow \ket{1,1}$ rotational transition to generate the dressed states $\ket{+}$ and $\ket{-}$. (b) The potential energy curves for two molecules plus the microwave field without (dashed) and with (solid) the microwave coupling for collisional angle $\theta = 90^\circ$.  The detunings are $\Delta = 2\pi\times 10~\rm{MHz}$ and, for the latter, $\Omega$ is  $2\pi\times 5~\rm{MHz}$.   
    } 
\label{fig1}
\end{figure*}

In this work, we realize the blue-detuned microwave loss suppression of ground-state bosonic \NaRb molecules and, with its help, demonstrate evaporative cooling for increasing the phase-space density (PSD). \NaRb molecules are non-reactive. However, in previous experiments, rapid two-body losses were still observed and attributed to the formation of $\rm Na_2Rb_2$ complexes at the short range~\cite{ye2018collisions, gersema2021probing}. As collisions between identical bosons proceed via the barrierless $s$-wave, the measured loss rate coefficients $\beta_{\rm in}$ of several times $10^{-10}~\rm{cm^3/s}$ are very large. As depicted in Fig.~\ref{fig1}, applying a circular polarized microwave blue detuned to the $J=0 \leftrightarrow J =1$ rotational transition can induce a long-range potential barrier, thus prevent molecules from reaching the short range and forming the complex~\cite{karman2019imperfectly, karman2020microwave, lassabliere2018controlling}. Here, we successfully suppress $\beta_{\rm in}$ to as low as $3.0(3)\times 10^{-12}~\rm{cm^3/s}$. This has allowed us to extract the elastic collision rate coefficient $\beta_{\rm el}$ from the cross-dimensional rethermalization and confirms the large good-to-bad collision ratio which is enough for an efficient evaporation at sample densities away from the hydrodynamic regime.

As illustrated in Fig.~\ref{fig1}(a), the rotational transition frequency between $\ket{J=0,m_J =0}$ and $\ket{1,1}$ for ground-state \NaRb is 4.180 GHz \cite{guo2018high}. Here, $J$ and $m_J$ are the rotational quantum number and its projection along the magnetic field, respectively. The AC electric field of a $\sigma^+$ polarized microwave signal near this frequency can interact with the $d_0 = 3.2$ Debye permanent dipole moment strongly. This polarizes the molecule rotating in the plane of the electric field and induces a time-average effective dipole moment $d_{\rm eff} = d_0/\sqrt{12(1+\Delta^2/\Omega^2)}$~\cite{yan2020resonant, karman2022resonant, deng2022effective}, with $\Omega$ the on-resonance Rabi frequency and $\Delta$ the detuning. However, the direct effective DDI alone, which is $\propto -d_{\rm eff}^2(1-3\cos^2\theta)/r^3$, cannot provide the 3D repulsive barrier between two molecules as it is attractive for collisional angles $\theta$ (w.r.t the quantization axis along the vertical direction) large than $54.7^\circ$. The second-order DDI, the microwave dressing modified rotational van der Waals interaction, must be included to ensure a repulsive barrier for all $\theta$. We note that because of the different signs of the dipolar interactions, the 3D long-range barrier exists only for blue-detuned circularly polarized microwave fields.    

%The effective two-molecule potential can be approximated as~\cite{deng2022effective}
%\begin{equation}
%   	V_{eff} = \frac{C_3}{r_3}P_2(\cos\theta)+\frac{C_6}{r^6}[7-5P_2(\cos\theta))-4P_4(\cos\theta)], 
%\label{eq:effDDI}
%\end{equation}
%which, as shown in by the red solid curve in Fig.~\ref{fig1}(b), has a barrier at around 800$a_0$ even for \textcolor{blue}{ $\theta=90^\circ$}.

The rigorous NaRb-NaRb adiabatic interaction potentials with microwave coupling can be derived with the multi-channel dressed state framework. In the simplified picture as illustrated in Fig.~\ref{fig1}(b), we first consider the bare two-molecule states $\ket{0,0}\ket{0,0}$, $\ket{0,0}\ket{1,1}$, $\ket{0,0}\ket{1,0}$, and $\ket{0,0}\ket{1,-1}$. Because of the resonant DDI between $J=0$ and $J=1$, the potential energy curves of the latter three states are all split into an upper repulsive branch and a lower attractive branch. In the dressed state picture without the coupling, the potential energy curve of $\ket{0,0}\ket{0,0}$ plus the photon energy crosses the repulsive branch, with the point of intersection moving toward short-range at larger blue detunings. In the presence of $\sigma^+$ polarized microwave, the \NaRb molecule is in the dressed states $\ket{+} = \sin{\phi}\ket{0,0}+\cos{\phi}\ket{1,1}$ and $\ket{-} = \cos{\phi}\ket{0,0}-\sin{\phi}\ket{1,1}$, with $\phi = {\rm arctan}(-\Omega/\Delta)/2$ the mixing angle. The microwave also causes coupling between the two-molecule state $\ket{0,0}\ket{0,0}$ and $\ket{0,0}\ket{1,1}$ and leads to an avoided crossing near the Condon point $R_c = (d_0^2/12\pi\epsilon_0 \hbar\tilde{\Omega})^{1/3}$~\cite{karman2022resonant}, with $\tilde{\Omega}=\sqrt{\Omega^2+\Delta^2}$ the generalized Rabi frequency. The potential energy curve of the resulting two-molecule dressed state $\ket{++}$ has a rather high barrier for inter-molecular distance $R<R_c$. For the parameters used in Fig.~\ref{fig1}(b), the calculated $R_c$ locates near 680$a_0$ which is far away from the inter-molecular range for forming the $\rm Na_2Rb_2$ complex. Two-body losses due to complex formation can therefore be effectively suppressed, at the expense of some small amount of microwave-induced losses~\cite{karman2018microwave,lassabliere2018controlling}. In addition, the DDI also induces a large elastic collision cross section. With the loss suppressed and the elastic collisions enhanced, this opens the possibility of evaporative cooling of \NaRb.

%$\ket{+} = \sin{\phi}\ket{0,0}+\cos{\phi}\ket{1,1}$ and $\ket{-} = \cos{\phi}\ket{0,0}-\sin{\phi}\ket{1,1}$, with $\phi = \arctan(-\Omega/\Delta)/2$ the mixing angle.

%\section{Experimental condition}

\begin{figure}[t]
    \centering
    \includegraphics[width=0.8\linewidth]{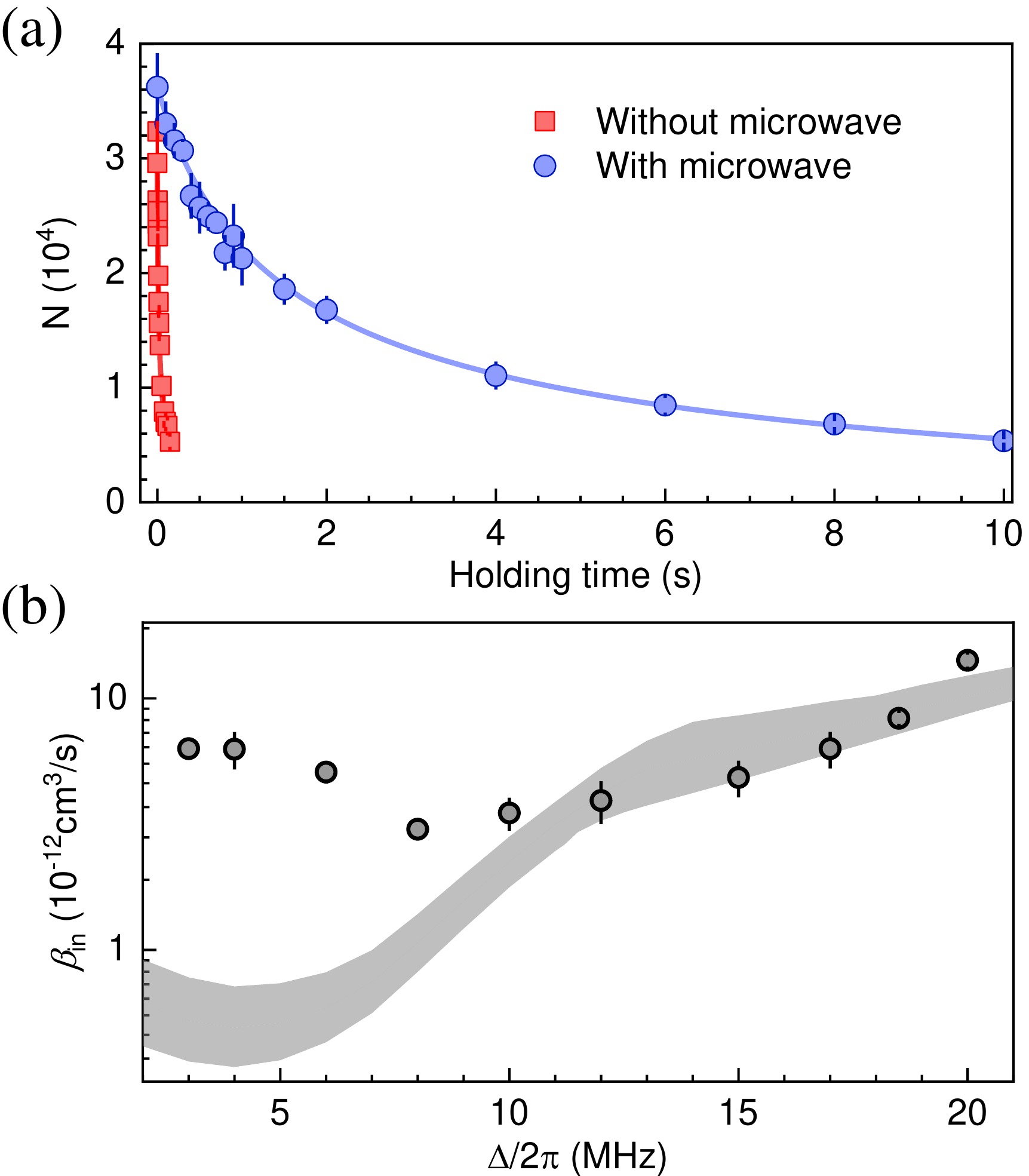}
\caption{
Loss suppression. (a) The decay of \NaRb number $N$  without (red squares) and with (blue circles) the microwave field. For the latter case, the detuning $\Delta$ is $2\pi\times8~\rm{MHz}$. The solid curves are for eye guiding. Error bars represent standard deviation of typically 6 shots. (b) $\beta_{\rm in}$ versus $\Delta$ for temperatures of about 700 nK. The shaded band is from the coupled-channel calculation for $\xi$ between $5^\circ$ and $7^\circ$. Error bars are from fitting.      }
    \label{fig2}
\end{figure}

\section{Experiment configurations}

The experiment starts from optically trapped samples of ultracold ground-state \NaRb molecules prepared in the nuclear spin stretched state of the $\ket{0, 0}$ rotational level via magneto-association and stimulated Raman population transfer~\cite{guo2016creation}~(Appendix~\ref{GSMcreate}). The oscillation frequencies of the molecules in the crossed optical dipole trap are $(\omega_x,\omega_z,\omega_y)=2\pi\times(40,40,200)~\rm{s^{-1}}$. For typical samples with $3.5\times 10^4$ molecules and a temperature $T$ of 700 nK, the calculated peak density reaches $n = 4.6\times10^{11}~\rm{cm^{-3}}$. At this temperature, the measured $\beta_{\rm in}$ without any loss suppression is $3.8(5)\times 10^{-10}~\rm{cm^3/s}$. As shown in the inset of Fig.~\ref{fig2}(a) the two-body loss limited effective lifetime $1/\beta_{\rm in} \expval{n}$, with $\expval{n}=n/\sqrt{8}$ the mean density, is only 16 ms.

The microwave signal is generated by a low phase noise signal generator and amplified to about 50 W by a high power amplifier~(Appendix~\ref{microwaveSetup}). To further suppress noises with sideband frequencies near $\Omega$, which could drive non-adiabatic transitions between the dressed states and thus sabotage the shielding effect, we add two 10 MHz bandwidth bandpass filters in series before the amplifier. The amplified signal is then split into two paths and fed into a doubly-fed waveguide antenna. The relative phase and amplitude between the two paths are then fine-tuned to achieve the purest $\sigma^+$ polarization. From Rabi oscillations of the $\sigma^+$, $\sigma^-$, and $\pi$ rotational transition driven with low microwave powers, the best measured polarization ratios are $E_{\sigma^+}: E_{\sigma^-}: E_{\pi} = 1: 0.054(2): 0.072(3)$, corresponding to an ellipticity angle $\xi$ of $3.1(3)^\circ$. For the rest of the work, unless otherwise specified, the on-resonance Rabi frequency is fixed to $\Omega = 2\pi \times 9.8(3) $ MHz~(Appendix~\ref{microwaveRabi}).

\section{Loss suppression}

 To observe the loss suppression, we need to prepare the $\ket{J=0, m_J=0}$ molecules in the $\ket{+}$ dressed state adiabatically. This is done by ramping the microwave power up exponentially in $100~\rm{\mu s}$ right after the Raman transfer. After some holding time $t$, we apply a reversed microwave ramp to convert the molecules back to the bare state for detection. Measurements from the back-to-back microwave ramping, with very short holding time in between, show that the molecular dressed state is prepared with near perfect fidelity. Figure~\ref{fig2}(a) shows a comparison of the molecule number versus the holding time without and with the microwave. A dramatic increase of the lifetime from less than 20 ms to over 1 s is observed, indicating microwave shielding for \NaRb is working.

However, to extract the two-body $\beta_{\rm in}$ from the data, several other effects must be taken into account. The residual microwave noise~\cite{chen2012general}, as well as off-resonance photon scattering from the trapping light~\cite{lin2022seconds, vexiau2017dynamic}, can cause one-body losses while evaporation following elastic collisions leads to extra two-body losses. With low density samples with $\expval{n} \approx 2\times 10^{10}~ \rm{cm^{-3}}$, we minimize loss contributions from two-body processes and measure a total one-body loss rate of $\Gamma = 0.119(3) ~\rm s^{-1}$ in microwave fields at $\Delta = 2\pi \times 8$ MHz. The photon scattering induced one-body loss rate calculated from the measured trapping light intensity is $\gamma_{\rm L} = 0.107(4)~\rm s^{-1}$. Assuming $\Gamma = \sqrt{\gamma_{\rm L}^2 + \gamma_{\rm MW}^2}$, the microwave noise induced one-body loss rate $\gamma_{\rm MW}$ should be $0.052(11)~\rm s^{-1}$. Based on the model of the phase noise induced loss, the calculated phase noise near the Rabi frequency is -159.5(9) dBc/Hz~\cite{chen2012general}. To minimize two-body losses from evaporation, we raise the trap depth $U$ to increase the truncation factor $\eta=U/k_B T$ to larger than 9. Here $k_B$ is the Boltzmann constant. With the smaller Boltzmann factor $e^{-\eta}$, this reduces the two-body loss contribution from evaporation to below 9\% which, for simplicity, we choose to neglect~(Appendix~\ref{Fullmodel}).

We then measure the number evolution versus holding time $t$ in the dressed state for a range of $\Delta$. We fit the data to the rate equation~\cite{ni2010dipolar}
\begin{equation}
    \frac{dn}{dt}= -\beta_{\rm{in}} n^2 - \Gamma n
\label{eq:twobody}
\end{equation}
to extract $\beta_{\rm in}$ for each $\Delta$ with $\Gamma$ obtained independently for the same microwave configurations following the aforementioned one-body loss measurement~(Appendix~\ref{onebodyloss}). Only data at short $t$ are used to keep the temperature increase from the anti-evaporation effect to less than 30\% so that we can consider $\beta_{\rm in}$ a constant in the evolution of molecule number. The peak density $n$ is calculated from the measured trap frequencies and the sample temperature.

The measured $\beta_{\rm in}$ are summarized in Fig.~\ref{fig2}(b) with the lowest value of $3.0(3)\times 10^{-12}~\rm{cm^3/s}$ occurring at $\Delta = 2\pi\times8~\rm{MHz}$. The loss is suppressed by a factor of 125 compared to the value of $3.8(5)\times 10^{-10}~\rm{cm^3/s}$ measured with samples of the same initial conditions without the microwave shielding. Away from the minimum point, $\beta_{\rm in}$ becomes larger in both directions. The increase of $\beta_{\rm in}$ at larger $\Delta$ can be understood intuitively by the decrease of $R_c$ and thus less effective loss suppression. However, the $\beta_{\rm in}$ increase for smaller $\Delta$ cannot be understood following this picture. 

This discrepancy is also observed when comparing the experiment with the coupled-channel calculation. The shaded band in Fig.~\ref{fig2}(b) represents the calculated $\beta_{\rm in}$ for $\xi$ between $5^\circ$ and $7^\circ$. This range of $\xi$, which is slightly larger than the measured value from the low power calibration, is chosen for the best matching with the measured $\beta_{\rm in}$ values for $\Delta \geq \Omega$. However, for small detunings, the disagreement between theory and experiment is obvious. Qualitatively, the calculated $\beta_{\rm in}$ also starts to increase for very small $\Delta$, but the turning point occurs around  $2\pi \times 4$ MHz, much smaller than the observed one. More importantly, even with the large polarization imperfection, the calculated $\beta_{\rm in}$ can still reach well below $10^{-13}~\rm cm^3/s$, much smaller than the best measured value. Currently, we cannot tell whether this disagreement is of technical origin, for instance the detuning dependent phase noise due to the finite bandwidth of the filters, or from some unknown loss processes not accounted for.

\section{Elastic collision}

\begin{figure}[tbh]
\centering
  \includegraphics[width=0.9\linewidth]{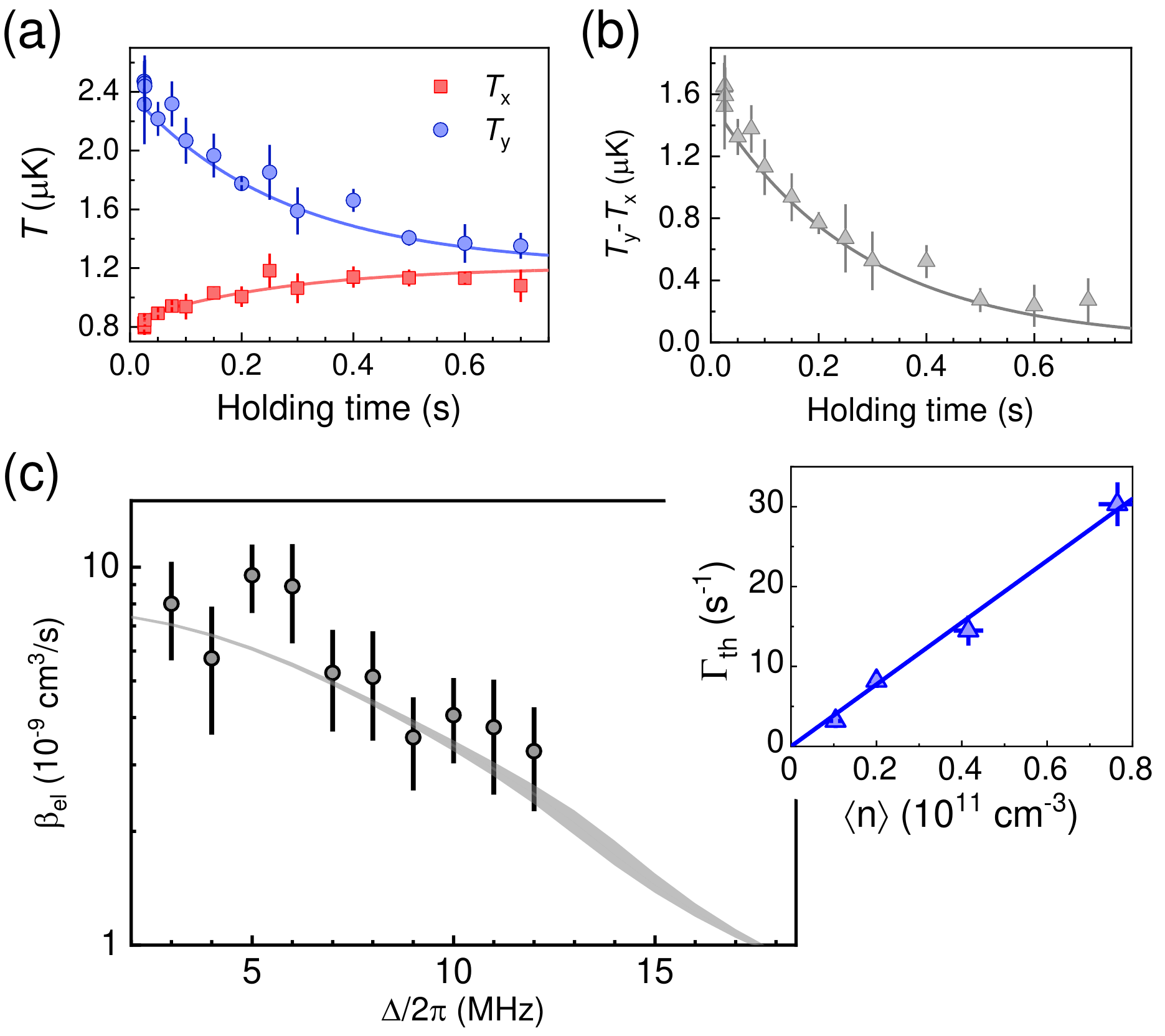}
\caption{
Measurement of $\beta_{\rm el}$. (a) Example cross-dimensional rethermalization data after the parametric heating along the y-direction for $\Delta = 2\pi\times11~\rm{MHz}$. $T_x$ and $T_y$ are the temperatures along the horizontal and the vertical directions, respectively. (b) shows the temperature difference $T_x-T_y$ for extracting the thermalization rate $\Gamma_{\rm th}$. Solid curves are from exponential fitting. (c) $\beta_{\rm el}$ versus $\Delta$. The shaded band is from the coupled-channel calculation for $\xi$ between $5^\circ$ and $7^\circ$. The inset shows the linear dependence of $\Gamma_{\rm th}$ on the mean density $\expval{n}$ at $\Delta = 2\pi\times 8~\rm{MHz}$. The error bars in (a) and (b) represent the standard deviation, while those in (c) are from fitting.
    }
\label{fig3}
\end{figure}

Next we measure $\beta_{\rm el}$ with the cross-dimensional rethermalization method in the presence of microwave shielding. To this end, as shown in Fig.~\ref{fig3}(a), we heat the sample along y direction with parametric heating and measure the evolution of $T_x$ and $T_y$. We fit the measured temperature difference $T_y - T_x$ with the exponential function to extract the thermalization rate $\Gamma_{\rm th}$ [Fig.~\ref{fig3}(b)]. $\beta_{\rm el}$ is then calculated from $\Gamma_{\rm th} = \expval{n} \beta_{\rm el}/ N_{\rm col}$~\cite{tang2015swave, patscheider2022determination, li2021tuning, chen2023field}, with $N_{\rm col}$ the number of collisions per rethermalization. In our system, $N_{\rm col}$ ranges from 10 to 16 due to the strong DDI~\cite{wang2021thermalization}~(see Appendix~\ref{NcolCal}). This is much larger than the typical value of 2.5 for pure $s$-wave collisions.

As $\beta_{\rm el}$ is expected to be very large, the hydrodynamic limit, in which $\Gamma_{\rm th}$ is capped to the trap frequency, may be reached with our highest density samples. This will make the measured $\beta_{\rm el}$ smaller than the real value. To avoid this, the densities are reduced by first holding the samples for different durations before the parametric heating. As can be seen from the inset of Fig.~\ref{fig3}(c), the linear dependence of $\Gamma_{\rm th}$ on density proves that the hydrodynamic limit is not reached for the typical densities used in this measurement. The lower densities also ensure that the number loss and heating from the inelastic collisions are small and thus need not to be considered here.  

Figure \ref{fig3}(c) shows the measured $\beta_{\rm el}$ for $\Delta$ from $2\pi\times 3$ MHz to $2\pi\times 12$ MHz using this method. Beyond this range, the reduced $\beta_{\rm el}/\beta_{\rm in}$ ratio makes the measurement different due to the longer thermalization time and the non-negligible number loss and heating from lossy collisions. Indeed, the $\beta_{\rm el}$ values are very large, reaching the level of $10^{-8}~\rm{cm^3/s}$ for smaller detunings and thus larger $d_{\rm eff}$. The shaded band is the theoretical calculation for $\xi$ between $5^\circ$ and $7^\circ$, same as the range of $\xi$ used in Fig.~\ref{fig2}(b). Unlike $\beta_{\rm in}$, $\beta_{\rm el}$ shows almost no dependence on $\xi$. A good agreement between experiment and theory is clearly visible.

\section{Evaporative cooling} 

\begin{figure}[t]
    \centering
    \includegraphics[width=0.9\linewidth]{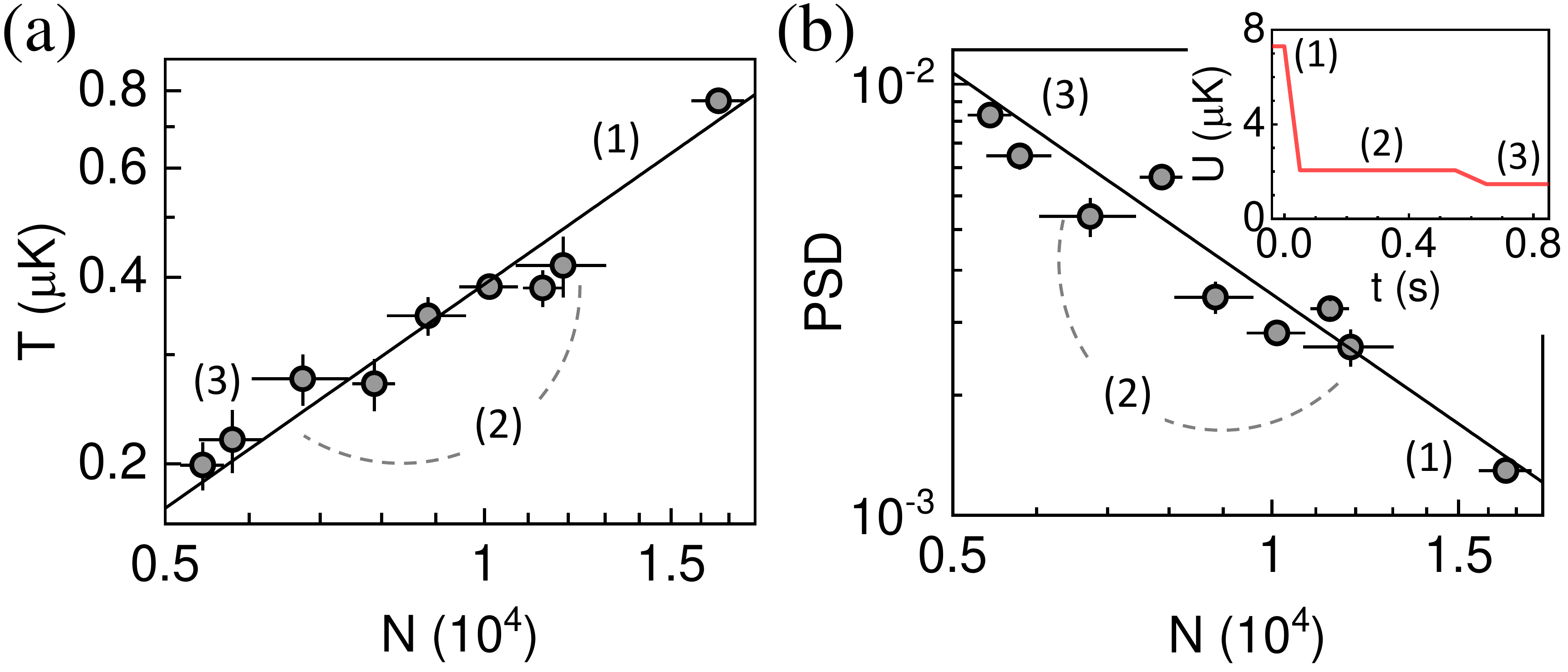}
    \caption{
    Evaporative cooling of \NaRb with microwave shielding.
	(a) and (b) are the trajectories of the temperature and the PSD at different molecule numbers in log-log scale, respectively. The solid lines are linear fits for extracting the evaporation efficiencies. The inset of (b) depicts the decrease in trap depth $U$ with evaporation time $t$. Evaporation mainly occurs in the constant power regions between the two ramp steps, as marked by the three numbers. The error bars for $T$ are from fitting while for $N$ and PSD represent the standard deviation.
    }
    \label{fig4}
\end{figure}

At $\Delta = 2\pi \times 8~\rm{MHz}$ and temperature $T=800~\rm{nK}$, the ratio between the elastic collision rate $\expval{n}\beta_{\rm el}$ and the loss rate $\expval{n}\beta_{\rm in}$, the so called good-to-bad collision ratio, can reach up to 1500. This should support evaporative cooling with high efficiency. Indeed, as shown in Fig.~\ref{fig4}, by decreasing the optical trap depth in two steps we observe decrease of the temperature from 800 nK to 200 nK and increase of the PSD from 0.001 to 0.01 after losing the number of molecules from 16000 to 5000. The temperature decrease following a slope $\dv*{\ln{T}}{\ln{N}} = 1.2(1)$ while the evaporation efficiency $-\dv*{\ln{\rm{PSD}}}{\ln{N}}$ is 1.7(2). To our knowledge, this is the highest evaporation efficiency ever achieved for ultracold molecules.

Rapid loss is observed when the trap depth is further reduced, most likely due to the disappearance of the trapping potential along the gravitational direction. We have tried to further improve the evaporation by introducing an additional trapping beam from the vertical direction to tighten the confinement. However, this only results in an evaporation efficiency smaller than one. We attribute this to the higher density which brings the system into the hydrodynamic limit~\cite{ma2003evaporative}. 
As the thermalization rate is capped while the loss rate keeps increasing, the good-to-bad collision ratio becomes worse and the evaporation becomes less efficient. For the measurement in Fig.~\ref{fig4}, the peak density stays around $1.6\times 10^{11}~\rm{cm^{-3}}$ during the whole process and the hydrodynamic limit is never reached. However, while high efficiency evaporative cooling is feasible in this parameter range, to achieve a BEC, which requires higher density, $\beta_{\rm in}$ needs to be lowered further.

\section{Conclusion} 

To summarize, we establish microwave shielding as a convenient yet powerful way for suppressing the loss of bosonic UPMs to enable efficient evaporative cooling. Although the BEC of UPMs has not yet been achieved, results demonstrated here are already useful for many interesting experiments. For instance, loading the sample into optical lattices following the evaporation and with the microwave shielding, the filling factor should be increased significantly so that effects of DDI can be probed directly~\cite{lin2022seconds}. The large $\beta_{\rm el}$ makes it possible to enter the hydrodynamic regime with currently achievable densities. Interesting behaviors of thermal gases with strong DDIs have been predicted in this regime~\cite{wang2022thermoviscous}.

This work represents an important step toward the creation of a BEC of bosonic UPMs. Future studies will be focused on ways to lower $\beta_{\rm in}$ further so that high good-to-bad collision ratios can be maintained in the high density regime. Following our theoretical predictions, $\beta_{\rm in}$ on the low $10^{-13} ~\rm cm^3/s$ level should be achievable with higher microwave Rabi frequencies. The shielding effect can also be enhanced by using a quasi-1D potential to constrain the collision angle $\theta$ to small values. Eventually, these improvements will engineer favorable collision properties for creating the BEC of UPMs which, once achieved, will lead to the realization of many of the exotic quantum phases with strong and long-range DDIs~\cite{micheli2006toolbox,yi2007,gorshkov2011quantum,baranov2012condensed,schmidt2022self}.

While preparing this manuscript, we became aware of similar work on microwave shielding with bosonic NaCs molecules~\cite{bigagli2023}. We obtained hints regarding the calculation of $N_{\rm col}$ from their results.

\begin{acknowledgments}
We are grateful to X. Y. Luo, J. L. Bohn and T. Karman for valuable discussions, and Bo Yang for laboratory assistance. We thank K. L. Wu and Y. M. Fung for lending us a signal generator at the early stage of the project. The Hong Kong team was supported by Hong Kong RGC General Research Fund (Grants No. 14301818, No. 14301119, and No. 14302722) and Collaborative Research Fund (Grants No. C6009-20GF). W. Z. acknowledges support from the National Natural Science Foundation of China (NSFC) under Grants No.12274331. T. S. and S. Y. acknowledge support from the NSFC (Grants No. 12135018, and No. 11974363).

\end{acknowledgments}

\appendix

\section{Microwave system}
\label{MWsystem}
\subsection{Generating the high power and low noise microwave signal}
\label{microwaveSetup}

\begin{figure}
    \centering
    \includegraphics[scale=0.38]{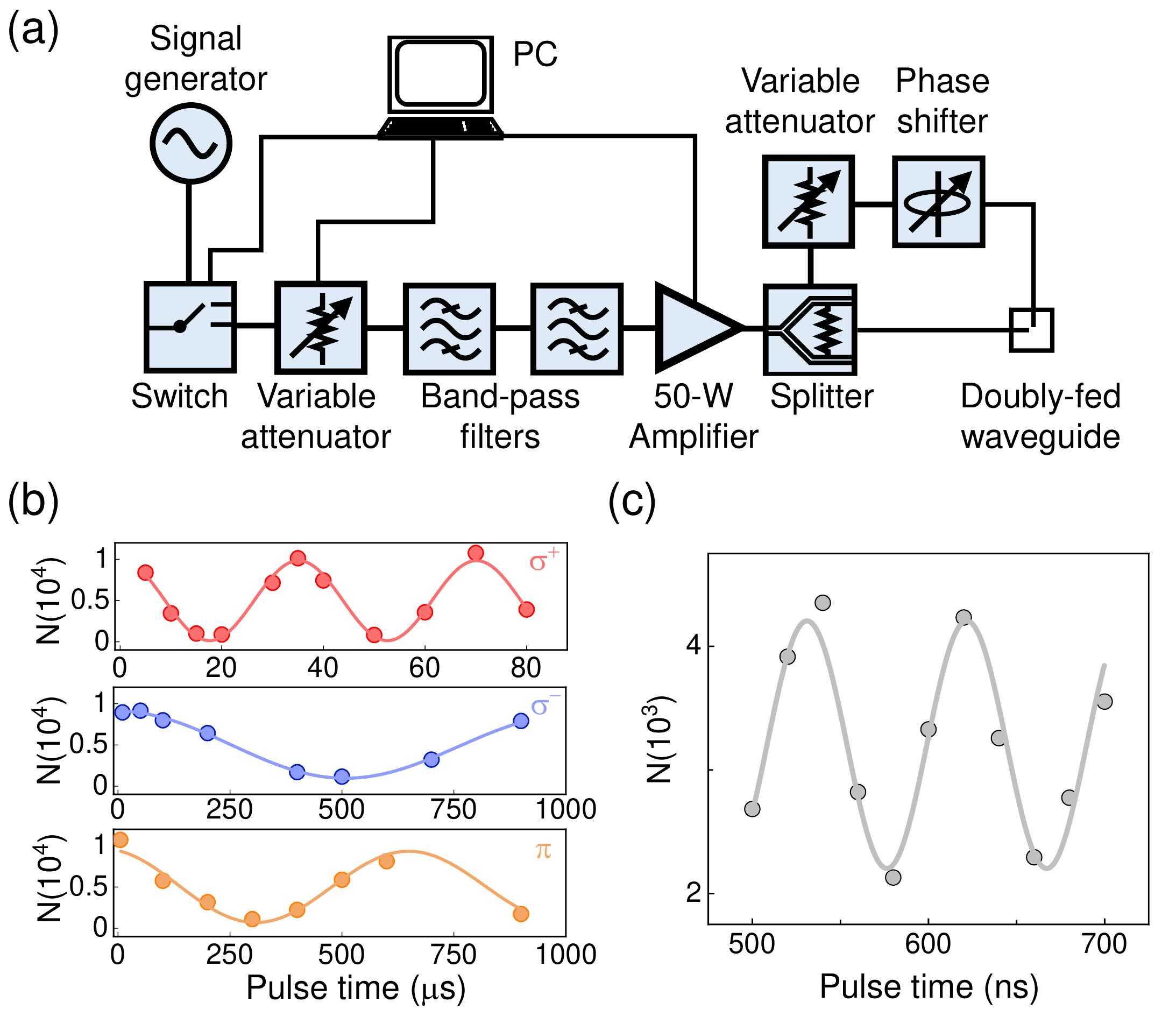}
    \caption{(a) Schematic of the microwave system. The passive components after the amplifier are all rated to high powers. (b) Rabi oscillations of the $\sigma^{+}$, $\sigma^{-}$, and $\pi$ transitions, i.e, the $\ket{0,0}\leftrightarrow \ket{1,1}$, $\ket{0,0}\leftrightarrow \ket{1,-1}$, and $\ket{0,0}\leftrightarrow \ket{1,0}$ for microwave polarization calibration. (c) The detuned Rabi oscillation for measuring $\Omega$ at the high microwave power used in this work. The $\Delta$ used is $2\pi\times5 ~\mathrm{MHz}$. The measured on resonance Rabi frequency is $\Omega = 2\pi\times9.8(3) ~\mathrm{MHz}$.}
    \label{figS1}
\end{figure}

As shown schematically in Fig.~\ref{figS1}, we use a low noise signal generator (HP/Agilent 8665B) as the 4.2 GHz microwave source. The signal then passes through a switch followed by a variable attenuator which is used for the adiabatic creation of the dressed state. The combined maximum isolation of the switch and the attenuator is about -120 dB. The phase noises for sidebands near the Rabi frequency are further suppressed by two bandpass filters with 10 MHz bandwidths before the signal is amplified to approximately 50 W by a low noise figure high-power amplifier (Microwave Amps AM43). The amplified signal is then split into two paths and fed into the two input ports of a doubly-fed waveguide. The waveguide is fabricated by soldering five pieces of printed circuit boards together with dimensions of $40.8~\rm{mm}\times 40.8~\rm{mm}\times 68.3~\rm{mm}$. A mechanical phase shifter and a mechanical variable attenuator, both rated to high powers, are placed before one of the input ports of the waveguide for relative phase and power balance control to generate an output signal with near perfect circular polarization~\cite{chen2023field}.

In our experiment, the bandpass filters are playing a very critical role. Without the filters, the signal generator's phase noise is not good enough to achieve the current long one-body lifetime and the two orders of magnitude suppression of loss. However, because of the fixed center frequency and the finite bandwidth of the filters, the phase noise of the filtered signal at sidebands near the Rabi frequency changes with the carrier frequency. Also, the phase noises for sidebands within the passband of the filters are not suppressed. The residual $\sigma^-$ and $\pi$ polarized signal output from the waveguide may cause accidental near-resonant dressing with other nuclear spin levels. These issues have not been taken into account in the theoretical modelling. In future works, these issues can be mitigated by using a signal generator with better low frequency phase noise or by adding a high quality factor microwave cavity as an additional narrow bandwidth filter.

\subsection{Polarization and Rabi frequency calibrations}
\label{microwaveRabi}
To calibrate the microwave polarization, the low power Rabi oscillations between the rotational excited states $\ket{1, 1}$, $\ket{1, -1}$, $\ket{1, 0}$ and the ground-state $\ket{0, 0}$ at 335 G are measured as shown in Fig.~\ref{figS1}(b). The power of the microwave field is low enough so that off-resonance coupling can be ignored. The optical trap is turned off during the microwave pulse to avoid the differential ac Stark shift. The measured ratios of electric field components are $E_{\sigma^{+}}: E_{\sigma^{-}}: E_{\pi} = 1:0.054(2):0.072(3)$. These correspond to a polarization ellipticity angle $\xi$ in the frame of the microwave field of $3.1(3)^\circ$ and a tilting angle of the microwave propagation direction with respect to the magnetic field of $5.6(3)^{\circ}$. 
%Here, the error bars mainly come from the unknown relative phase between the $\pi$-component and circular components. 
The microwave polarization is assumed to be the same for high power operation.

The Rabi frequency used in the microwave shielding experiment is measured with detuned Rabi oscillations. As the $E_{\sigma^{-}}$ and  $E_{\pi}$ components are small, at large detunings, their contributions can be safely neglected. Fig.~\ref{figS2}(c) shows the Rabi oscillation of the $\ket{0,0}\leftrightarrow \ket{1,1}$ transition with detuning $\Delta = 2\pi\times5~\rm{MHz}$. From the generalized Rabi frequency $\tilde{\Omega}$ obtained from the fit, the on-resonance Rabi frequency $\Omega=\sqrt{\tilde{\Omega}^2-\Delta^2}$ is $2\pi\times9.8(3)~\rm{MHz}$.

\section{Sample preparation and detection}
\label{GSMcreateAnddetect}
\subsection{Creation of ground-state \NaRb}
\label{GSMcreate}
Detailed procedures for preparing ground-state \NaRb have been discussed in our previous works~\cite{guo2016creation}. In this work, we have made some adjustments to reduce the loss caused by atom-molecule collisions. Briefly, immediately after the magneto-association with the Feshbach resonance at 347.65 G, we apply a stimulated Raman adiabatic passage (STIRAP) at 346 G to transfer the population to the rovibrational ground state $\ket {J=0, m_J=0, m^{\rm{Na}}_{I}=3/2, m^{\rm{Rb}}_{I}=3/2}$. Here, $m^{\rm{Na}}_{I}$ and $m^{\rm{Rb}}_{I}$ are projections of the \Na~ and \Rb~ nuclear spins, with $I= 3/2$ for both, to the quantization axis as defined by the B-field. As only the nuclear spin stretched states are used in this work, for simplicity, we use $\ket {J, m_J}$ to label the single molecule states throughout this work. 
%The STIRAP takes 50 $\rm \mu s$ and throughout the process the trap lights are off. 
Residual Na and Rb atoms are then quickly removed with resonant light in less than 1 ms without affecting the ground-state molecules. Afterwards, the molecular sample is further purified by pulsing on a magnetic field gradient. 

As densities of the residual atoms are high, atom-molecule collisions induced losses are one of the main factors limiting the number of molecules in the group state. Compared with the previous procedures with the STIRAP and atom removal take place after ramping the magnetic field to 335 G, this new procedure shortens the interaction time between the residual atoms and molecules substantially. As a result, a 60\% increase in molecule numbers is routinely obtained.

\subsection{Detection and thermometry}
\label{GSMDetect}

For detection, the ground-state molecules have to be transferred back to Feshbach molecules. Starting from the dressed state, we first ramp down the microwave power in 100 $\mu$s to bring the molecules back to the bare $\ket{0,0}$ state. The optical dipole trap is then turned off abruptly and the reversed STIRAP sequence is applied to transfer the population to the weakly bounded state. After some time-of-flight (TOF) expansion time $t$, the molecule number and RMS size $\sigma(t)$ of the sample are determined from the high-field absorption images of the Rb atoms right after breaking down the \NaRb Feshbach molecules to free Na and Rb atoms with a 10 $\mu$s photodissociation pulse~\cite{jia2020detection}.

Typically, the temperature $T$ is determined from fit $\sigma(t)$ to $\sigma(t) = \sqrt{\sigma^2_0+k_B T t^2/m}$ with the initial size $\sigma_0$ and $T$ as the free parameters. Here $m$ is the mass of $^{23}$Na$^{87}$Rb. For the cross-dimensional rethermalization experiment, long expansion time is not possible because of the high temperature caused by the parametric heating. Instead, we use a fixed expansion time, typically $t = 1.4$ ms, and determine the temperature from $\sigma(t) = \sqrt{k_B T/(m \omega^2_{x,y})  + k_B T t^2/m}$ with the measured trap frequencies $\omega_{x}$ and $\omega_{y}$. With lower temperature samples, we have measured that the temperature differences obtained from these two methods are less than 14\%.

\section{Modeling the loss}
\label{ModeltheLoss}
\subsection{The full model}
\label{Fullmodel}
As mentioned in the main text, multiple one- and two-body processes can lead to number losses and temperature changes. To measure $\beta_{\rm in}$, contributions from all other processes must be accounted for. Including all these processes, the number and temperature evolution can be modelled by the coupled differential equations  
\begin{align}
\dot{N}&=- \beta_{\rm ev}A \frac{N^2}{T^{3/2}} -\beta_{\rm in}A \frac{N^2}{T^{3/2}}-\Gamma N, \label{eqns:number}\\
\begin{split}
\dot{T}&=\frac{3-\eta-\kappa}{3} \beta_{\rm ev}A \frac{N}{T^{1/2}}+(\frac{1}{4}+h_0)\beta_{\rm in}A \frac{N}{T^{1/2}}, \label{eqns:temp}
\end{split}
\end{align}
where $A=(\Bar{\omega}^2m/4\pi k_{B})^{3/2}$ with $\Bar{\omega}=(\omega_x \omega_y \omega_z)^{1/3}$ the mean trap frequency. The $\beta_{\rm ev}$ terms account for the number loss and cooling from evaporation induced by elastic collisions. In the deep trap limit with truncation factors $\eta \geqslant 6$, it can be approximated as $\beta_{\rm ev}\approx(\eta-4)e^{-\eta}\beta_{\rm el}/N_{\rm col}$ with $\kappa\approx(\eta-5)/(\eta-4)$~\cite{olson2013optimizing,luiten1996evaporation}. $(\eta+\kappa)k_B T$ is the averaged energy  carried away by an evaporated particle. The $\beta_{\rm in}$ terms are the number loss and heating from two-body inelastic collisions, including contributions from both complex formation and microwave induced loss~\cite{karman2018microwave,lassabliere2018controlling}. We include the $h_0$ term to account for heating from all additional possible contributions~\cite{ye2018collisions}. The $\Gamma$ term describes the total one-body losses which include contributions from the microwave phase noise~\cite{anderegg2021observation,schindewolf2022evaporation} and off-resonance scattering of the trapping light. Collisions with background gas molecules can also cause one-body losses, but for our UHV system the time scale of these collisions is too long to be relevant. 

In practice, it is very difficult, if not impossible, to reliably extract all the one- and two-body loss coefficients from a single number and temperature evolution measurement. Therefore, we choose to measure them sequentially, taking advantage of their very different dependencies on density and trap depth.

\subsection{The one-body loss}
\label{onebodyloss}
\begin{figure}
    \centering
    \includegraphics[scale=0.42]{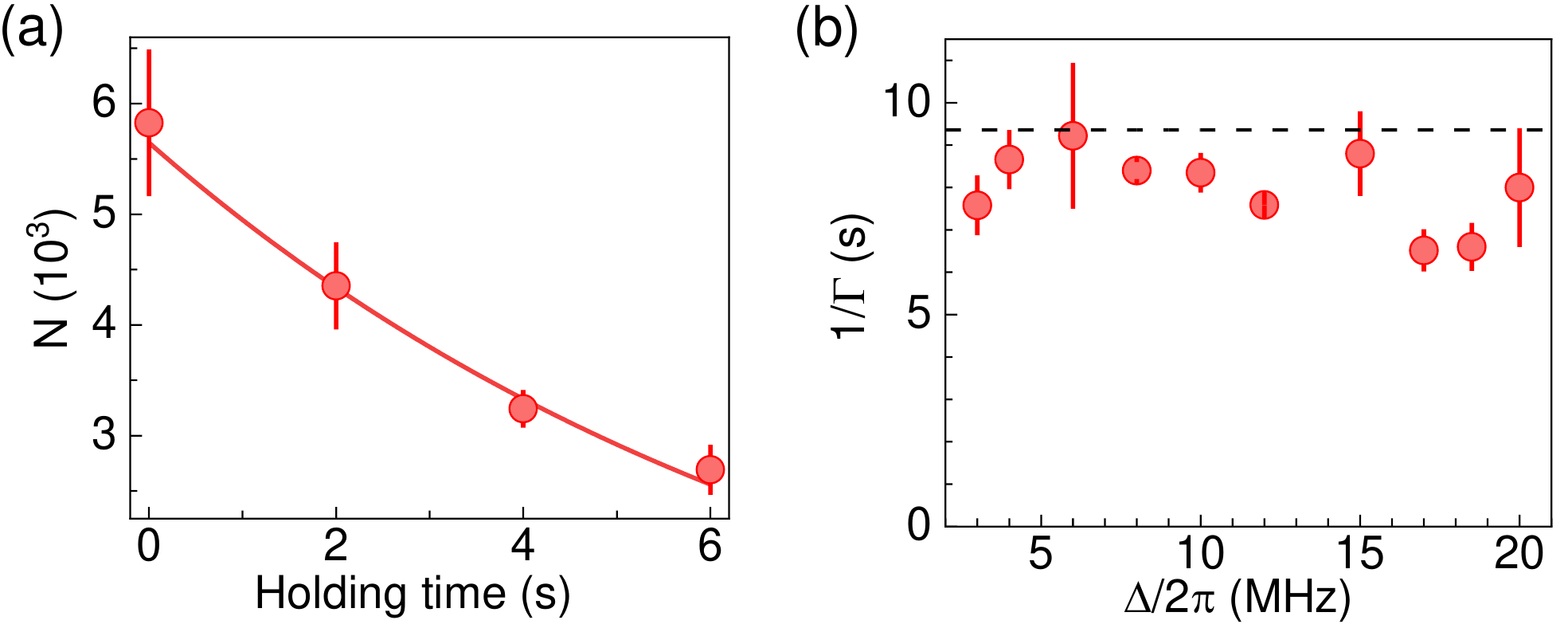}
    \caption{(a) Number evolution at $\Delta=2\pi\times3~\rm MHz$ for low density samples. $\Gamma$ is extracted by exponential fitting as shown by the solid curve. (b) The one-body lifetime $1/\Gamma$ versus $\Delta$. The dashed line is the lifetime limited by the off-resonance scattering of the optical trapping light.}
    \label{figS2}
\end{figure}

As only two-body processes are density dependent, when the density is low enough, their contributions to the loss will become less important compared with those from one-body processes. As shown in Fig.~\ref{figS2}(a), we control the density by reducing the molecule number until the observed loss is dominated by one-body processes. From the number evolution, the total $\Gamma$ is obtained by exponential fitting. This is done for all microwave detunings involved, as summarized in Fig.~\ref{figS2}(b). 

Figure~\ref{figS2}(b) also shows the loss rate $\gamma_{\rm L}$ induced by scattering of the trapping light which is calculated from the measured light intensity and the imaginary polarizability. From the total loss rate $\Gamma = \sqrt{\gamma_{\rm L}^2 + \gamma_{\rm MW}^2}$, the one-body loss induced by the microwave phase noise $\gamma_{\rm MW}$ is then determined. The microwave phase noise at the Rabi frequency is calculated from $\gamma_{\rm MW}$ with the microwave phase noise induced one-body loss model~\cite{avdeenkov2006suppression}.

\subsection{$\beta_{\rm in}$ measurement}
\label{twobodylossMeasure}
\begin{figure}
    \centering
    \includegraphics[scale=0.42]{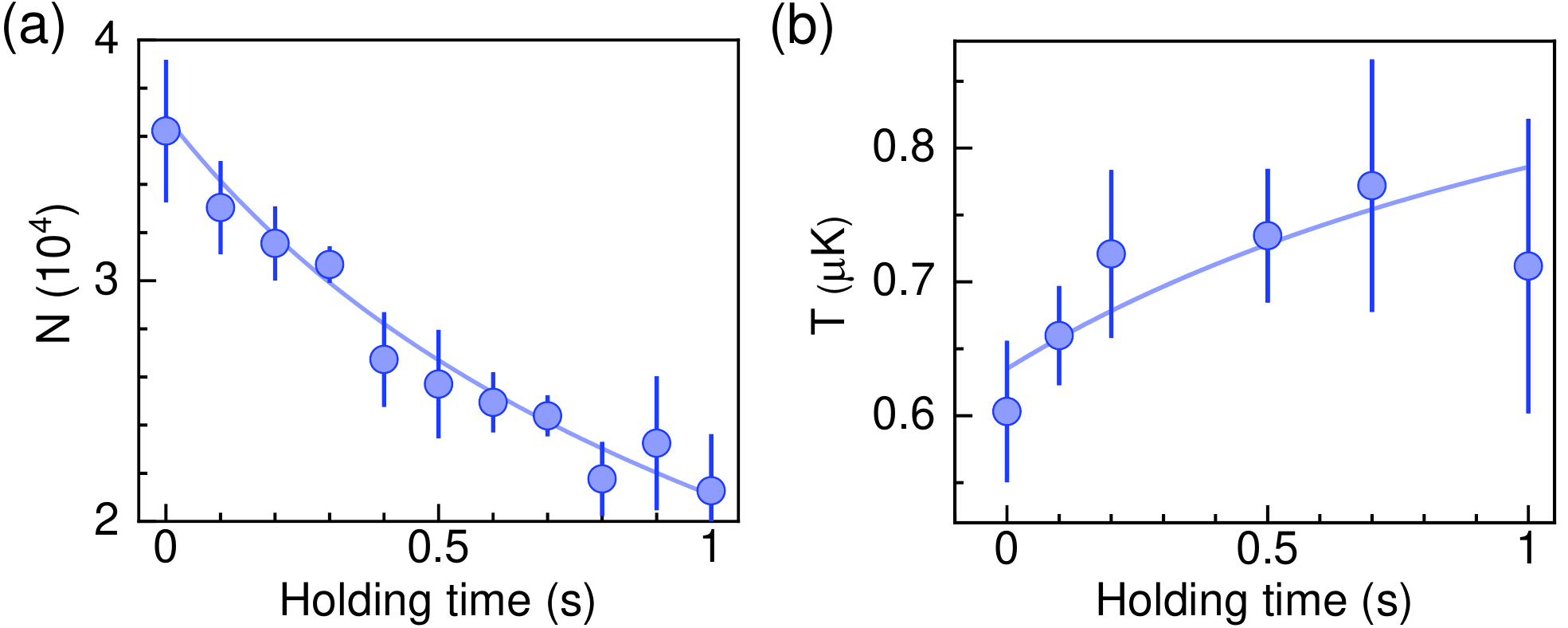}
    \caption{(a) and (b) Number and temperature evolution at $\Delta=8~\rm MHz$. The blue solid lines are from the joint fitting following Ref.~\cite{ye2018collisions}. See text for details. }
    \label{figS3}
\end{figure}

With $\Gamma$ measured independently, we further simplify the model by using $\eta \geq 9$. In such a deep trap, $\beta_{\rm ev}$ is less than 9\% of $\beta_{\rm in}$ because of the $(\eta-4)e^{-\eta}$ factor, despite the very large $\beta_{\rm el}$. The number loss from evaporation is then neglected in the analysis for extracting $\beta_{\rm in}$. We also only use data at short holding times to ensure the temperature increase is less than 30\% [Fig.~\ref{figS3}(b)]. We then ignore the temperature evolution and use only Eq.~\ref{eqns:number} to extract $\beta_{\rm in}$ from the number evolution. These simplifications lead to the model in the main text
\begin{equation}
    \frac{dn}{dt}= -\beta_{\rm{in}} n^2 - \Gamma n.
\label{eq:twobodyS}
\end{equation}
In the density calculation, the mean temperature during the number evolution is used. 

We note that the temperature evolution can also be included in the simplified model. $\beta_{\rm in}$ can then be extracted from the joint fit to the number and temperature data, as done in our previous work~\cite{ye2018collisions}. As demonstrated in Fig.~\ref{figS3}, for $\Delta=2\pi\times\rm 8\ MHz$, we obtain $\beta_{\rm in}=3.3(3)\times10^{-12}\ \rm cm^3/s$ from this joint fitting. This is within the error bar of the $3.0(2)\times10^{-12}\ \rm cm^3/s$ value obtained from fitting the number evolution with Eq.~\ref{eq:twobodyS} only. 

An important issue revealed by the joint number and temperature fitting is the large additional heating. The extracted $h_{\rm 0}$ with the microwave shielding is 0.36(0.17). This is larger than that of the bare state which is 0.12(6). Currently, the mechanism behind this surprising observation is still unclear. The microwave induced loss is indeed accompanied by energy release on the order of $\hbar \Omega$. This is much larger than the trap depth and secondary collisions with other molecules, which are required for transferring energy to the sample, seem unlikely. However, we do not know whether there are other off-resonance transitions, for instance, driven by the phase noise, which could produce trappable hot molecules to deposit energy to the sample.

\section{Theoretical calculations}
\label{theory}
\subsection{Calculations of $\beta_{\rm el}$ and $\beta_{\rm in}$}
\label{multichannelCal}
We calculate the elastic scattering rate and the two-body inelastic loss rate by closely following the approach detailed in Ref.~\cite{deng2022effective}. Here we briefly outline the main procedure for the numerical calculations.

The NaRb molecules are treated as rigid rotors with electric dipole moments. For a single molecule, only the lowest two rotational levels with total four rotational states are considered in our calculation. As a result, the two-molecule Hilbert space is of sixteen dimensions. To achieve microwave shielding, all molecules are prepared in the highest dressed state, $|+\rangle$, in an elliptically polarized microwave field. Consequently, we only need to consider the scattering of two identical molecules in the ten-dimensional Hilbert subspace consisting of the symmetrized two-body states. Fortunately, it turns out that there are only seven of these two-molecular states are mutually coupled, which defines the scattering channels, $\nu=1$ to $7$, in the multichannel calculations. In our convention, the $\nu=1$ channel corresponds to the two-molecule state $|++\rangle$ and the incident energy $E$ is defined with respect to the asymptotical energy of $\nu=1$ channel.

To proceed, we expand the wavefunction of the $\nu$th channel in the partial-wave basis as $\psi_{\nu l m}$, where $l$ is up to $L_{\rm max}=8$. In numerical calculations, we impose a capture boundary condition~\cite{clary1987chemical} at $r=50\,a_0$ from which the scattering wavefunctions are numerically propagated to a large distance $r_M$ ($>5\times 10^4\,a_0$) using the log-derivative method~\cite{Johnsom1973log-derivative}. Then by comparing the wavefunctions with the asymptotic boundary condition, we obtain the scattering $K$-matrix which subsequently leads to the scattering $S$-matrix with element being denoted as $S_{\nu lm}^{\nu'l'm'}$. The elastic and inelastic scattering cross sections can be computed according to
\begin{align}
    \sigma_\mathrm{el}(E)&=\frac{2\pi}{k^2}\sum_{lml'm'}\left|S_{1 lm}^{1l'm'}-\delta_{ll'}\delta_{mm'}\right|^2,\\ \sigma_\mathrm{in}(E)&=\frac{2\pi}{k^2}\sum_{lml'm'}\left(\delta_{ll'}\delta_{mm'}-\left|S_{1 lm}^{1l'm'}\right|^2\right),
\end{align}
where $k=\sqrt{mE/\hbar^2}$ is the incident momentum with $m$ being the mass of the molecule. Finally, to calculate $\beta_\mathrm{el}$ and $\beta_\mathrm{in}$, we assume a Maxwell-Boltzmann distribution for the thermal gases of temperature $T$. Then the elastic scattering and the two-body inelastic loss rates are~\cite{yan2020resonant}
\begin{align}
    \beta_\mathrm{el,in}
    =\sqrt\frac{16k_BT}{\pi m}\frac{1}{(k_BT)^2}\int_0^\infty \sigma_\mathrm{el,in}(E)  e^{-\frac{E}{k_BT}}E dE,\label{betainel}
\end{align}
where $k_B$ is the Boltzmann constant. The integral in Eq.~\eqref{betainel} is numerically computed with $45$ incident energies logarithmically spaced between $0.02\,k_BT$ and $8\,k_BT$. The computed $\beta_\mathrm{el}$ and $\beta_\mathrm{in}$ are compared to experimental measurements in the main text.

\subsection{Dipolar thermalization}
\label{NcolCal}
\begin{figure}
    \centering
    \includegraphics[scale=0.9]{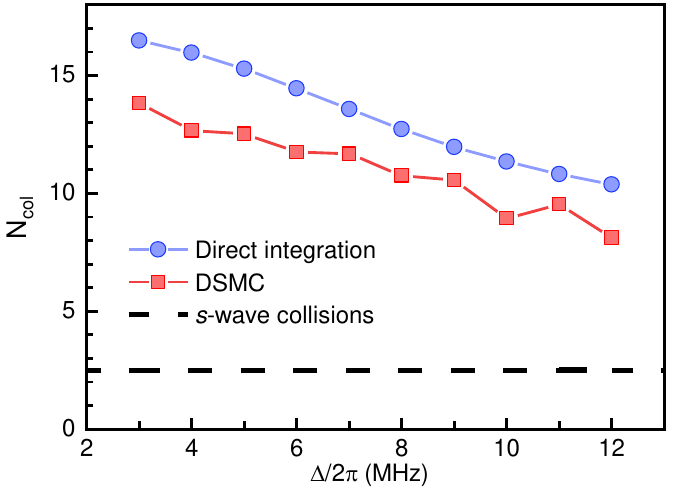}
    \caption{The number of collisions per rethermalization obtained from the direct numerical integration (blue circles) and the DSMC simulation (red squares). The dashed line shows $N_{\rm col}=2.5$ for \textit{s}-wave collisions.}
    \label{figS4}
\end{figure}

To measure $\beta_{\rm el}$ in the cross-dimensional re-thermalization experiment, it is necessary to determine the number of collisions required for each thermalization, denoted as $N_{\rm col}$, in order to calculate $\beta_{\rm el}$ from the measured thermalization rate $\Gamma_{\rm th}$ using the formula $\Gamma_{\rm th} = \expval{n} \beta_{\rm el}/ N_{\rm col}$. To this end, we begin by computing the collision integral in the Enskog equations~\cite{wang2021thermalization} 
\begin{align}
    \mathcal{C}[\Delta p_j^2] = \int dq^3 \frac{n^2(q)}{N}\int \frac{dp_r^3}{2m} p_r c_r(p_r)\int d\Omega_{p'} \frac{d\sigma}{d\Omega_{p'}} \Delta p_j^2, \label{eqcollisioninte}
\end{align}
using the differential cross-section ${d\sigma}/{d\Omega_{p}}$ obtained from the aforementioned multichannel calculation. Here, $j \in \{x,y,z\}$ denotes the corresponding Cartesian coordinate in the laboratory frame, $n^2(q)$ refers to the spatial distribution, and $c_r(p_r)$ represents the distribution of relative momenta $p_r$.

We then express the rethermalization rate as
\begin{align}
    \Gamma_{\rm th} =\frac{\mathcal{C}[\chi_j]}{\expval{\chi_j}} = \frac{\mathcal{C}[\Delta p_j^2]}{2m \expval{\chi_j}},
\end{align}
with $\expval{\chi_j} = k_B(\mathcal{T}_j-T_{\rm eq})$ the phase-space averaged quantity in which $\mathcal{T}_j = m \omega_j^2 \expval{q_j^2}/2 k_B + \expval{p_j^2}/2m k_B$ is the pseudo-temperature, and $T_{\rm eq}$ is the final equilibrium temperature. Finally, $N_{\rm col}$ is calculated directly using the equation
\begin{align}
    N_{\rm col} = \frac{\expval{n}\expval{\sigma_{\rm el}\upsilon}}{\Gamma_{\rm th}},
    \label{eqNcol}
\end{align}
with $\expval{n}$ the mean density, and $\expval{\sigma_{\rm el}\upsilon}$ the thermal averaged cross section.

In Fig.~\ref{figS4}, the blue circles are $N_{\rm col}$  for different $\Delta$ calculated following this procedure. As depicted by the dashed line, we can observe that $N_{\rm col}$ is significantly larger than that for \textit{s}-wave collisions, which is only 2.5. This is due to the highly anisotropic nature of the microwave-dressed potential~\cite{deng2022effective}. $N_{\rm col}$ is smaller for larger $\Delta$ because of the reduction in the effective dipole moment $d_{\rm eff}$. Nevertheless, even for the largest $\Delta$ used in the rethermalization measurement, $N_{\rm col}$ is still above 10. 

To cross-check our calculations, we also simulate the thermalization using the direct-simulation Monte Carlo (DSMC) method~\cite{Sykes2015Nonequilibrium}, employing stochastic sampling based on the differential cross section to determine the post-collision momenta. As shown in Fig.~\ref{figS4}, The DSMC simulations confirm the validity of our calculations, with the obtained $N_{\rm col}$ values approximately 15\% lower than those from direct calculations but with a consistent overall trend. 

Finally, the $\beta_{\rm el}$ values presented in Fig.~\ref{fig3}(c) of the main text are calculated using the $N_{\rm col}$ values obtained from the direct numerical integration.

%\bibliography{MWrefs} 
%

\end{document}